\documentclass[showpacs,superscriptaddress,nofootinbib,preprintnumbers,amsmath,amssymb]{revtex4}
\usepackage{graphicx}
\usepackage{dcolumn}
\usepackage{bm}
\usepackage{subfigure}

\begin{document}

\preprint{}

\title{The impact of dissipation and noise on fluctuations in chiral fluid dynamics}

\author{Marlene Nahrgang}
\affiliation{SUBATECH, UMR 6457, Universit\'e de Nantes, Ecole des Mines de Nantes,
IN2P3/CRNS. 4 rue Alfred Kastler, 44307 Nantes cedex 3, France }
\affiliation{Frankfurt Institute for Advanced Studies (FIAS), Ruth-Moufang-Str.~1, 60438 Frankfurt am Main, Germany}

\author{Christoph Herold}
\affiliation{Frankfurt Institute for Advanced Studies (FIAS), Ruth-Moufang-Str.~1, 60438 Frankfurt am Main, Germany}
 \affiliation{Institut f\"ur Theoretische Physik, Goethe-Universit\"at, Max-von-Laue-Str.~1,
 60438 Frankfurt am Main, Germany}

  \author{Stefan Leupold}
 \affiliation{Department of Physics and Astronomy, Uppsala University, Box 516, 75120 Uppsala, Sweden}

  \author{Igor Mishustin}
 \affiliation{Frankfurt Institute for Advanced Studies (FIAS), Ruth-Moufang-Str.~1, 60438 Frankfurt am Main, Germany}

\author{Marcus Bleicher}
\affiliation{Frankfurt Institute for Advanced Studies (FIAS), Ruth-Moufang-Str.~1, 60438 Frankfurt am Main, Germany}
 \affiliation{Institut f\"ur Theoretische Physik, Goethe-Universit\"at, Max-von-Laue-Str.~1,
 60438 Frankfurt am Main, Germany}

\date{\today}

\begin{abstract}

We investigate the nonequilibrium evolution of the sigma field coupled to a fluid dynamic expansion of a hot fireball to model the chiral phase transition in heavy-ion collisions. The dissipative processes and fluctuations are allowed under the assumption that the total energy of the coupled system is conserved. We use the linear sigma model with constituent quarks to investigate the effects of the chiral phase transition on the equilibration and excitation of the sigma modes. The quark fluid acts as a heat bath in local thermal equilibrium and the sigma field evolves according to a semiclassical stochastic Langevin equation of motion. The effects of supercooling and reheating of the fluid in a first order phase transition are observed via the delayed relaxation of the sigma field to a new equilibrium state. Nonequilibrium fluctuations of the sigma field in a scenario with a first order phase transition are much stronger than in a critical point scenario.

\end{abstract}

\pacs{25.75.−q, 47.75.+f, 11.30.Qc, 24.60.Ky}

\maketitle

\section{Introduction}

At high  temperatures and densities strongly interacting matter is supposed to form a state called the quark-gluon plasma, which can experimentally be created and studied in relativistic heavy-ion collisions. A broad variety of experimental signatures is proposed to study the QCD phase diagram in the laboratory. These signatures cover different aspects of the deconfinement and the chiral phase transition, including a suggested critical point \cite{Stephanov:1998dy,Stephanov:1999zu} and the first order phase transition \cite{Mishustin:1998eq,Randrup:2010ax}.
It is, however, not clear how much of a potential signal of the phase transition is developed in a realistic scenario of a heavy-ion collision. The fast dynamics during the creation and expansion of the fireball of strongly interacting matter makes it difficult to defer results from fluid dynamic calculations directly. A thorough theoretical understanding of phase transitions in the environment of a heavy-ion collision is necessary to make profound predictions for upcoming experiments at the low energy RHIC program \cite{Caines:2009yu}, the CBM project at FAIR \cite{Friman:2011zz} and at NICA \cite{nica:whitepaper} experiments. At a critical point the relaxation time for an order parameter becomes infinite. Therefore, a system that cools rapidly through a second order phase transition is necessarily driven out of equilibrium and any signal for a critical point is weakened \cite{Berdnikov:1999ph}. A fast expansion, however, leads to interesting phenomena at a first order phase transition, such as supercooling \cite{Csernai:1992tj} and spinodal decomposition \cite{Chomaz:2003dz}. An intersting signal was proposed for a nonequilibrium chiral phase transition, i.e. the enhancement of soft pions from the decay of disoriented chiral condensates (DCC) \cite{Rajagopal:1993ah,Biro:1997va,Xu:1999aq}.
The theoretical description of such nonequilibrium processes can be achieved by a Langevin equation. It has been derived explicitly within a nonequilibrium quantum field formalism for the $\phi^4$ model \cite{Morikawa:1986rp,Gleiser:1993ea,Boyanovsky:1996xx,Greiner:1996dx}, gauge theories \cite{Bodeker:1995pp,Son:1997qj} and ${\cal O}(N)$ chiral models \cite{Rischke:1998qy}. Phenomenological treatments of stochastic Langevin equations for the propagation of the chiral fields have been performed to study the formation of disoriented chiral condensates \cite{Biro:1997va} or to study hadron production in a spinodal decomposition scenario \cite{Fraga:2004hp}. These studies were done at either a fixed temperature or assuming a scaling expansion and cooling of the system.

In the following we address the two questions: how much of the equilibrium behavior at a critical point survives in a dynamical environment and how much of nonequilibrium effects at a first order phase transition survive relaxational dynamics?

Our calculations below are performed within a chiral fluid dynamic model, where the propagation of chiral fields is coupled to a fluid dynamical evolution of the medium. In \cite{Mishustin:1998yc} this coupled dynamics was desribed within the linear sigma model with constituent quarks. In this model, the fluid consists of quarks and antiquarks which are considered in local thermal equilibrium and propagated according to energy-momentum conservation of an ideal fluid. Alternatively, in \cite{Csernai:1995zn,Mishustin:1997ff,Abada:1996bw} the quarks were described by a Vlasov equation in the collisionless approximation. In \cite{Paech:2003fe,workinprogress} initial fluctuations of the sigma field were introduced and propagated deterministically during the evolution of the system. Then large oscillations of the chiral fields were observed, which did not show relaxational behavior of the order parameter. In the present work we go beyond deterministic equations of motion for the chiral fields by including dissipation and noise in the framework of a Langevin equation. As already mentioned above we will utilize the linear sigma model with constituent quarks for this purpose.

This paper is organized as follows: in section \ref{sec:chiralfluiddyn} we present the coupled model of extended chiral fluid dynamics starting from the linear sigma model with constituent quarks and in section \ref{sec:numerics} we explain details of its numerical implementation with special emphasis on the energy-momentum conservation during the evolution of the entire system.
 In section \ref{sec:cf2_supercooling} we investigate the nonequilibrium evolution of the coupled dynamics leading to supercooling, reheating and an enhanced intensity of the nonequilibrium fluctuations of the sigma field. Section \ref{sec:summ} is devoted to summary and outlook.

\section{Extended chiral fluid dynamics}\label{sec:chiralfluiddyn}

\subsection{General remarks}
 Our work is an extension of existing chiral dynamic models \cite{Greiner:1996dx,Biro:1997va,Rischke:1998qy,Mishustin:1998yc,Paech:2003fe} including relaxational dynamics of the order parameter of the chiral phase transition in order to study the effects of nonequilibrium fluctuations. Due to the interaction with the quark fluid the chiral fields are damped. This is taken into account by friction terms in the Langevin equation of motion of the chiral fields. Due to these terms, the energy of the chiral field fluctuations dissipates into the quark fluid. To fulfil the dissipation-fluctuation theorem \cite{landaulifschitz5} the field fluctuations are included as a noise term on the right hand side of the Langevin equation, which reads
\begin{equation}
 \partial_\mu\partial^\mu\sigma+\frac{\delta U}{\delta\sigma}+\frac{\delta \Omega_{\bar qq}}{\delta\sigma}+\eta\partial_t \sigma=\xi\, .
\label{eq:equi_langevineq}
\end{equation}
It contains a classical Mexican hat potential $U$, the quark contribution to the thermodynamic potential $\Omega_{\bar qq}$ to one-loop level, the damping coefficient $\eta$ and the stochastic noise field $\xi$ \footnote{The last term in the l.h.s. of equation (\ref{eq:equi_langevineq}) is written in the rest frame of the fluid. To restore a covariant form of the equation one should replace $\partial_t\rightarrow u^\mu\partial_\mu\sigma$, where $u^\mu$ is the $4$--velocity of the fluid element.}. The dynamics of the quarks is reduced to the propagation of densities according to energy-momentum conservation, i.e. the equations of relativistic fluid dynamics
\begin{equation}
\partial_\mu T^{\mu\nu}=S^\nu\, ,
\label{eq:fluidT}
\end{equation}
where the source term $S^\nu$ accounts for the energy-momentum exchange between the fluid and the field.

 Our model is a combination and an extension of both approaches: the chiral fluid dynamics and the Langevin description of the chiral fields at the phase transition mentioned above. This model was derived in \cite{Nahrgang:2011mg} within the formalism of the two-particle irreducible (2PI) effective action \cite{Luttinger:1960ua,Lee:1960zza,Baym:1961zz,Baym:1962sx,Cornwall:1974vz,Ivanov:1998nv,vanHees:2001ik,vanHees:2001pf,vanHees:2002bv}, which gives the Langevin equation for the sigma field and the dynamics of the quarks consistently. The field and the fluid are coupled at different levels of the evolution of the entire system:
\begin{itemize}
 \item The one-loop \textbf{effective potential} $V_{\rm eff}=U+\Omega_{\bar qq}$ for the sigma field in presence of the quarks drives the chiral phase transition. Variation with respect to the sigma field gives the one-loop scalar density $\frac{\partial\Omega\bar qq}{\partial\sigma}=g\rho_s$, which is the standard mean-field contribution to the equation of motion (\ref{eq:equi_langevineq}) of the chiral field.
\item The pressure and the energy density of the locally equilibrated quark fluid in the \textbf{equation of state} $p(e)$ depend on the local value of the sigma field, which plays the role of an external parameter in the thermodynamic sense. 
  \item The \textbf{damping coefficient} $\eta$, which describes the dissipative and stochastic processes in the equation of motion of the sigma field, is given by the interaction with the heat bath. It depends on the temperature of the heat bath and consistently on the sigma field itself via the masses of the quarks and of the sigma mesons.
\item To account for the \textbf{energy-momentum exchange} between the two sectors we constructed an energy-momentum tensor, which is conserved for the full 2PI effective action. The approximations used for the derivation of the dynamics of the sigma field generate additional terms in the energy-momentum balance.
\end{itemize}
In the present work we implement this consistent coupling numerically and study the full nonequilibrium dynamics of the entire system with respect to effects of the phase transition. A detailed description of the theoretical aspects of the model can be found in  \cite{Nahrgang:2011mg}.

\subsection{The linear sigma model with constituent quarks}
We use the linear sigma model with constituent quarks as a low energy effective model of QCD.
It exhibits some important features of the suggested chiral phase structure of QCD. It has a crossover transition at vanishing $\mu$ and a first order transition line at high $\mu$ and lower temperatures, which terminates in a critical point at $T_c\simeq99$~MeV and $\mu_c\simeq207$~MeV \cite{Scavenius:2000qd}. The quantitative parameters of the phase transition in the linear sigma model may differ from the values for the true QCD. Therefore, we expect only a qualitative similarity to the true QCD phase transition.

The Lagrangian of the theory reads
\begin{equation}
\begin{split}
{\cal L}=&\overline{q}\left[i\gamma^\mu\partial_\mu-g\left(\sigma+i\gamma_5\tau\vec{\pi}\right)\right]q \\
  &+ \frac{1}{2}(\partial_\mu\sigma\partial^\mu\sigma)+
 \frac{1}{2}(\partial_\mu\vec{\pi}\partial^\mu\vec{\pi}) 
- U(\sigma, \vec{\pi}) \, .
\label{eq:LGML}
\end{split}
\end{equation}
Here $q=\left(u,d\right)$ is the constituent quark field, $\sigma$ and $\vec{\pi}$ are the sigma and the pion fields. The strength of the coupling between the quarks and the chiral fields is $g$. In the vertex for the pion-quark coupling the $\gamma_5$ matrix enters to account for the pseudoscalar nature of the $\pi$-mesons and the isospin matrix $\vec{\tau}$ accounts for the iso-vector nature of the pions. The interaction between the chiral fields is given by the potential 
\begin{equation}
U\left(\sigma, \vec{\pi}\right)=\frac{\lambda^2}{4}\left(\sigma^2+\vec{\pi}^2-\nu^2\right)^2-h_q\sigma-U_0\, .
\label{eq:Uchi}
\end{equation} 
Usually the parameters in (\ref{eq:Uchi}) are chosen such that the vacuum properties of the system are reproduced. Chiral symmetry is spontaneously broken in the vacuum and the sigma field acquires a finite expectation value of $\langle\sigma\rangle=f_\pi=93$~MeV while the pions have zero vacuum expectation value $\langle\vec\pi\rangle=0$. The explicit symmetry breaking term is $h_q\sigma=f_\pi m_\pi^2$ with $m_\pi=138$~MeV. Thus $\nu^2=f_\pi^2-m\pi^2/\lambda^2$. Choosing $\lambda^2=20$ yields a sigma mass $m_\sigma=\sqrt{2\lambda^2 f_\pi^2 + m_\pi^2}\approx 604$~MeV, which is within the mass range assumed for the sigma state in the meson spectrum. In order to have zero potential energy in the ground state $U_0=m_\pi^4/(4\lambda^2)-f_\pi^2 m_\pi^2$. At a coupling $g=3.3$ the constituent quark masses in vacuum are $m_q=306.9$~MeV.

In the following we will use for the parameters the values given above -- except for the coupling constant $g$. In principle, one can introduce a finite baryo-chemical potential and vary it to get either a crossover, critical point or first-order transition. For our qualitative study, however, we follow  \cite{Scavenius:2000bb,Aguiar:2003pp} and vary the coupling strength $g$ to consider the different possibilities.

The grand-canonical partition function of the quarks in presence of the chiral field is
\begin{equation}
 Z=\int{\cal D}\overline q{\cal D} q{\cal D}\sigma{\cal D}\vec\pi\exp\left[\int{\rm d}^4 x  {\cal L}\right]\; .
\end{equation}
In the mean-field approximation $Z$ can explicitly be calculated. The thermodynamic potential to one-loop level is given by
\begin{multline} 
V_{{\rm eff}}(\sigma, \vec{\pi},T)=-\frac{T}{V}\ln Z=U\left(\sigma, \vec{\pi}\right)+\Omega_{\bar qq}(T,\sigma, \vec{\pi})\\
              =U\left(\sigma, \vec{\pi}\right)-2d_q T \int\frac{{\rm d}^3p}{(2\pi)^3}\ln\left[1+\exp\left(-\frac{E_q}{T}\right)\right] \, ,
\end{multline}
where the degeneracy factor $d_q=12$ is the product of $N_f=2$ flavors, $N_c=3$ colors and two spin states. The energy of the quarks
\begin{equation}
  E_q=\sqrt{\vec p^2+m_q^2}\, ,
\end{equation}
depends on the chiral fields via the dynamically generated quark masses
\begin{equation}
 m_q^2=g^2(\sigma^2+\vec\pi^2)\, .
\label{eq:quarkmass}
\end{equation}
At $\mu=0$ the strength of the phase transition can be tuned by varying the strength of the coupling $g$.  For $g=3.3$ the transition is a crossover, while for $g=3.63$ the potential becomes flat at $T_c=139.88$~MeV as for a continuous transition (critical point). And for $g=5.5$ the phase transition is discontinuous (first order phase transition). Here, one finds two degenerate minima at $T_c=123.27$~MeV corresponding to the two coexisting phases in a first order phase transition.

For the fluid dynamical description of a phase transition we need an equation of state, which relates the pressure to the energy density. Since the sigma field is not necessarily at its equilibrium value the equation of state will depend explicitly on the local value of the sigma field $\sigma(x)$. This is different to standard chiral equations of state used in fluid dynamic calculations.
The pressure of the quarks is locally given by 
\begin{equation}
  p(\sigma, \vec{\pi},T)= -V_{\rm eff}(\sigma, \vec{\pi},T)+U(\sigma, \vec{\pi})\; .
\label{eq:lsm_eos1}
\end{equation}
The equation of state is completed by obtaining the local energy density from the thermodynamic relation
\begin{equation}
 e(\sigma, \vec{\pi},T)= T\frac{\partial p(\sigma, \vec{\pi},T)}{\partial T}-p(\sigma, \vec{\pi},T)\; .
\label{eq:lsm_eos2}
\end{equation}
In the linear sigma model the sigma field is the order parameter of the chiral phase transition. In the rest of this paper we will neglect the fluctuations of the pion field and set $\pi=\langle\pi\rangle=0$ for all times.

\subsection{Inclusion of dissipation and noise}

The damping coefficient $\eta$ in the Langevin equation for the sigma mean-field was calculated in \cite{Nahrgang:2011mg} for the linear sigma model with constituent quarks. Since we are mainly interested in the long-range fluctuations of the sigma modes we use the damping coefficient for the $|\vec{k}|=0$ mode in the Langevin equation. It is
\begin{equation}
  \eta=g^2\frac{d_q}{\pi}\left[1-2n_{\rm F}\left(\frac{m_\sigma}{2}\right)\right]\frac{1}{m_\sigma^2}\left(\frac{m_\sigma^2}{4}-m_q^2\right)^{3/2}\, .
\label{eq:dampingcoeff}
\end{equation}
 It depends on the local value of the sigma field via the sigma mass and the quark mass. We calculate these masses by using the local equilibrium value of the sigma field $\sigma_{\rm eq}$. 
It is obtained by minimizing the effective potential $V_{\rm eff}$
\begin{equation}
 \frac{\partial V_{\rm eff}}{\partial\sigma}\biggl|_{\sigma=\sigma_{\rm eq}}=0\quad\text{and}\quad\frac{\partial^2 V_{\rm eff}}{\partial^2\sigma}\biggl|_{\sigma=\sigma_{\rm eq}}>0\, .
\end{equation}
The sigma mass is given by the curvature of the effective potential at the minimum
\begin{equation}
 m_{\sigma_{\rm eq}}^2(T)=\frac{\partial^2V_{\rm eff}(T, \sigma)}{\partial\sigma^2}\biggl|_{\sigma=\sigma_{\rm eq}}\, .
\end{equation}
Alternatively to the use of the equilibrium value one could try to use the actual value of the sigma field for the evaluation of the quark mass in (\ref{eq:quarkmass}). For the determination of the sigma mass one presumably would have to study the (local) response of the sigma field to small variations on the basis of the equation of motion (\ref{eq:equi_langevineq}). At the critical point the equilibrium sigma mass becomes very small, but we expect to see fluctuations, which can locally change this effective mass. Concerning a first-order phase transition it is known that in nonequilibrium in the spinodal region the sigma mass square becomes negative. In principle, the explicit $\sigma$-dependence of $\eta = \eta(\sigma,T)$ could take these nonequilibrium effects into account. Then, several subtle questions need to be addressed, e.g., which ``mass'' should enter formula (\ref{eq:dampingcoeff}), if the effective mass becomes tachyonic. In the present work, however, we disregard these issues and use the local equilibrium value for the determination of the sigma and quark masses which in turn enter the damping rate (\ref{eq:dampingcoeff}).

\begin{figure}
  \centering
  \includegraphics[width=0.45\textwidth]{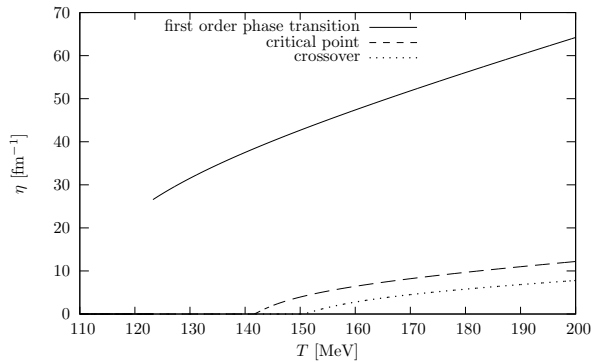}
  \caption{Temperature dependence of the damping coefficient $\eta$ for a the different couplings $g=5.5$, $g=3.63$ and $g=3.3$, which correspond to scenarios with a first order phase transition, a critical point and a crossover \cite{Nahrgang:2011mg}.}
  \label{fig:sc_dampfung}
 \end{figure}

This results in a temperature dependence of the damping coefficient $\eta$ as shown in figure \ref{fig:sc_dampfung}.
At lower temperatures it vanishes because the decay of the zero mode of the sigma field in a quark-antiquark pair is kinematically forbidden as $m_\sigma(T)<2m_q(T)=2g\sigma_{\rm eq}(T)$.
Albeit confinement is not included in the model, the vanishing of the damping process $\sigma \to \bar q q$ at low temperatures is in agreement with confinement. Physically the sigmas get further damped by interactions with the hard chiral modes \cite{Greiner:1996dx,Rischke:1998qy}, which are not included in our mean-field approach. To account for them we add to $\eta$ a value $2.2$/fm \cite{Biro:1997va}.
This contribution is attributed to the reactions $\sigma \leftrightarrow 2 \pi$. Consequently this damping effect should vanish when the sigma mass becomes smaller than $2m_\pi$. This occurs only in a critical point scenario, where the sigma mass is very small.
We note also that$\eta$ is very large, especially for the investigated scenario with a first order phase transition. This is due to the perturbative character of the derivation of $\eta$ and is manifested in the $g^2$ dependence of the result. This was discussed in \cite{Nahrgang:2011mg} and is considered to be improved in future work. 

To summarize, for $m_\sigma(T)>2m_q(T)=2g\sigma_{\rm eq}(T)$ the damping coefficient is given by equation (\ref{eq:dampingcoeff}). For $2m_q>m_\sigma(T)>2m_\pi$ we use the value $\eta=2.2/{\rm fm}$ and for $m_\sigma(T)<2m_\pi,2m_q$ $\eta=0$.
The stochastic field in the Langevin equation (\ref{eq:equi_langevineq}) has a vanishing expectation value
\begin{equation}
 \langle\xi(t)\rangle_\xi=0\, ,
\end{equation}
and the noise correlation is given by the dissipation-fluctuation theorem
\begin{equation}
 \langle\xi(t)\xi(t')\rangle_\xi=\frac{1}{V}\delta(t-t')m_\sigma\eta\coth\left(\frac{m_\sigma}{2T}\right)\, .
\label{eq:sc_noisecorrelation}
\end{equation}

\subsection{Source term for the quark fluid}
In \cite{Nahrgang:2011mg} we constructed a conserved energy-momentum tensor of the entire system using the formalism of the 2PI effective action. For any approximation to the full quark propagator the situation is more difficult due to the space-time dependence of the effective mass generated by the dynamic symmetry breaking. The total energy-momentum tensor can be represented as a sum of two contributions
\begin{equation}
T_{\rm total}^{\mu\nu}(x)=T_q^{\mu\nu}(x)+T_\sigma^{\mu\nu}(x)\, ,
\label{eq:sc_totaltmunu}
\end{equation}
where the first term comes from the quark-antiquark fluid and the second term from the sigma field.

The energy-momentum conservation equations, $\partial_\mu T^{\mu\nu}=0$ can be written as
\begin{equation}
 \partial_\mu T_q^{\mu\nu}(x)=-\partial_\mu T_\sigma^{\mu\nu}(x)=S^\nu(x)\, ,
\end{equation}
where $S^\nu$ is the source term for the fluid dynamics. In \cite{Nahrgang:2011mg} we have obtained the following approximate expression for the part of the sigma field

\begin{equation}
 \partial_\mu T_{\sigma, {\rm appr.}}^{\mu\nu}(x)=\left(-g\rho_s(x)-\eta(x)\partial_t\sigma(x)\right)\partial^\nu\sigma(x)\, .
\label{eq:sc_sourceterm}
\end{equation}
It consists of the standard mean-field result including the scalar density $\rho_s$ and the correction term given by the damping coefficient. It can, however, not account for the average energy transfer from the heat bath to the field given by the auxiliary noise field $\xi$. This issue was discussed in \cite{Nahrgang:2011ll}.

The quark contribution can be presented in symmetric form in terms of the quark propagator (dependent on center and relative variables $x$ and $u$ of the arguments)
\begin{equation}
 \partial_\mu T_{q,{\rm appr.}}^{\mu\nu}(x)\\
=-\frac{i}{2}\partial_\mu(\partial^\nu_u S_{\rm th}^{\scriptscriptstyle +-}(x,u)|_{u=0}\gamma^\mu+\partial^\mu_u S_{\rm th}^{\scriptscriptstyle +-}(x,u)|_{u=0}\gamma^\nu+...\\
\label{eq:sc_approxTmunu}
\end{equation}
The first term can be evaluated explicitly,
\begin{equation}
\begin{split}
 T_{q,{\rm th}}^{\mu\nu}(x)&=-\frac{i}{2}\left(\partial^\nu_u S_{\rm th}^{\scriptscriptstyle +-}(x,u)|_{u=0}\gamma^\mu+\partial^\mu_u S_{\rm th}^{\scriptscriptstyle +-}(x,u)|_{u=0}\gamma^\nu\right)\\
&=2 d_q\int\frac{{\rm d}^3p}{(2\pi)^3}\frac{p^\mu p^\nu}{p^0} n_{\rm F}(x,{\vec p})\, .
 \end{split}
\label{eq:sc_thermTmunu}
\end{equation}
It gives the energy-momentum tensor for an ideal fluid with the energy density and the pressure obtained from the equilibrium one-loop effective potential in mean-field approximation. This is exactly what we intend to use for the fluid dynamic description of the quark-antiquark fluid. Higher-order corrections to the energy-momentum tensor of the quark fluid are neglected in the present set up.
Then, the source term in equation (\ref{eq:fluidT}) can be identified with
 \begin{equation}
S^\nu=-\partial_\mu T_{\sigma, {\rm appr.}}^{\mu\nu}\, .
\label{eq:sourceterm}
 \end{equation}

\section{Numerical implementation}\label{sec:numerics}
\subsection{Fluid dynamics}

In the relativistic fluid dynamic equations (\ref{eq:fluidT})  the energy-momentum tensor of an ideal fluid reads
\begin{equation}
 T^{\mu\nu}=(e+p)u^\mu u^\nu - pg^{\mu\nu}\; ,
\label{hic_idealfluidtmunu}
\end{equation}
where $e$ is the local rest frame energy density, $p$ is the local rest frame pressure, $u^\mu=\gamma(1,\vec{v})$ is the local four-velocity of the fluid, $\gamma=(1-\vec{v}^2)^{-1/2}$ and $g^{\mu\nu}={\rm diag}(1,-1,-1,-1)$ the metric tensor. 

 For the solution of (\ref{eq:fluidT}) we use the full (3+1)d SHarp And Smooth Transport Algorithm (SHASTA) ideal fluid dynamic code \cite{Rischke:1995ir,Rischke:1995mt}. The causal transport of matter is assured on a numerical level by fulfilling the Courant-Friedrichs-Levy criterion $\Delta t/\Delta x\equiv \lambda<1$ and the SHASTA code requires values $\lambda<1/2$ for numerical stability. We use $\lambda=0.4$, $\Delta x=0.2$~fm and thus $\Delta t=0.08$. Our grid volume is $(128\Delta x)^3$. 

The numerical implementation of the source term (\ref{eq:sourceterm}) is as follows. 
After performing the fluid dynamic step in the standard fashion for $S^\nu=0$ we subtract the source term $S^\nu$ from the energy and momentum density in the computational frame. 
Then, we can again calculate the local rest frame quantities. This gives a very good conservation of energy and momentum of the entire system as we will show in section \ref{sec:cf2_energymomentumcons}. The latter method also proved to work well in multi-fluid dynamics \cite{Brachmann:1997bq}.

For the equation of state we do not obtain a simple relation $p(e)$ in the nonequilibrium evolution because the sigma field is not fixed at its equilibrium value. Therefore, the pressure (\ref{eq:lsm_eos1}) and the energy density (\ref{eq:lsm_eos2}) depend explicitly on the local value of the field, which can be viewed as an external parameter in the thermodynamic sense. Technically the following has to be done: With the energy density from the fluid dynamic calculation at a given point $x$, $e_{\rm fluid}(x)$ the equation for the thermodynamic energy density (\ref{eq:lsm_eos2}) needs to be inverted taking into account the local value of the sigma field $\sigma(x)$. The local temperature $T(x)$ is given by the solution of 
\begin{equation}
 e_{\rm fluid}(x)-e(\sigma,T)=0\, ,
\end{equation}
and used to calculate the thermodynamic pressure (\ref{eq:lsm_eos1}). For the transformations between the local rest frame of a fluid cell and the computational frame the equation of state is accessed very often in each time step. Since it is very time consuming to invert (\ref{eq:lsm_eos2}) numerically, we use a properly parametrized equation of state.

\subsection{Initial conditions}\label{sec:cf1_inicond}
For a qualitative investigation the initial conditions are kept simple. We choose an initial temperature profile, which is uniform in $z$-direction over a length $l_z=6$~fm and ellipsoidal in the $x/y$-plane. The ellipsoidal shape should mimic the overlap region in a heavy-ion collision. Its major and minor axes are $b=\sqrt{r_A^2-\tilde b^2/4}$ and $a=r_A-\tilde b/2$, where $\tilde b=6$~fm is the supposed impact parameter and $r_A=6.5$~fm the radius of Au nuclei. The temperature is smoothly distributed over this ellipsoidal region by a Wood-Saxon like distribution with the maximum at the initial temperature $T_{\rm ini}=160$~MeV, which is well above either of the phase transitions considered here,
\begin{equation}
 T(\vec x,t=0)=\frac{T_{\rm ini}}{(1+\exp((\tilde r-\tilde R)/\tilde a))(1+\exp\left((|z|-l_z)/\tilde a\right))}
\label{eq:lsm_iniT}
\end{equation}
with a surface thickness of $\tilde a=0.3$~fm, $\tilde r=\sqrt{x^2+y^2}$ and 
\begin{equation}
 \tilde R=\left\{
\begin{array}{rl}
\frac{ab\tilde r}{\sqrt{b^2x^2+a^2y^2}} & \mbox{for } \tilde r \neq 0 \\
a & \mbox{for }  \tilde r=0 \\
\end{array}
\right. \; .
\end{equation}
By minimizing the effective potential the equilibrium value of the sigma field $\sigma_{\rm eq}$ is found. The thermal equilibrium state has Gaussian fluctuations around the expectation value, the variance of which is
\begin{equation}
 \langle\delta\sigma^2\rangle=\frac{T}{V}\frac{1}{m_\sigma^2}\, .
\end{equation}
Thus, the sigma field is initially Gaussian distributed around its equilibrium value,
\begin{equation}
P\left(\sigma(\vec x)t=0\right)=\frac{1}{\sqrt{2\pi\langle\delta\sigma^2\rangle}}\exp\left[-\frac{(\sigma(x)-\sigma_{\rm eq}(x))^2}{2\langle\delta\sigma^2\rangle}\right]\, .
\label{eq:lsm_inisig}
\end{equation}
According to (\ref{eq:lsm_eos2}) the initial energy density of the quark fluid can be calculated with (\ref{eq:lsm_iniT}) and (\ref{eq:lsm_inisig}). 

The velocity profile is $v_z(\vec x, t=0)=|z|/l_z\cdot v_{\rm max}$, with $v_{\rm max}=0.2$, while initially there are no transverse velocities $v_x(\vec x, t=0)=v_y(\vec x, t=0)=0$.
Before we start the fluid dynamic expansion the equation of motion for the sigma field is solved a couple of times at $T_{\rm ini}$ to generate an initial distribution of the time derivative of the sigma field.
The pion field is initially set to zero and neglected during the simulation, since we expect that mainly the sigma field as the order parameter of chiral symmetry is affected by the phase transition.

\subsection{Energy-momentum conservation}\label{sec:cf2_energymomentumcons}
The energy-momentum conservation was discussed in \cite{Nahrgang:2011mg}, where we found a conserved energy-momentum tensor for the full 2PI effective action. In the truncated theory additional terms contribute to the energy-momentum balance. We first investigate the overall energy-momentum conservation before turning to the evolution of the system.

The correction to the divergence of the energy-momentum tensor of the sigma field in (\ref{eq:sc_sourceterm}) is given by the dissipative part of the equation of motion. It describes the energy transfer from the field to the fluid caused by the interaction between the fields and the fluid given by the pseudoscalar density $\rho_s$ and by dissipation for each numerical time step $\Delta t$
\begin{equation}
 \Delta E_\eta=\left(g\rho_s(x)+\eta(x)\partial_t\sigma(x)\right)\partial_t\sigma(x)\Delta t\, .
\end{equation}
In addition, the presence of the stochastic field $\xi$ causes an average energy transfer $\Delta E_{\xi}$ from the heat bath to the field. We determine $\Delta E_\xi$ by comparing $\Delta E_\eta$ to the numerically obtained energy difference in the sigma field before and after each time step of the simulation.

The total energy of the sigma field is given by
\begin{equation}
 E_{\sigma}=\frac{1}{2}(\partial_t\sigma)^2+\frac{1}{2}(\vec{\nabla}\sigma)^2+U(\sigma)\, .
\label{eq:cf_etotsigma}
\end{equation}
It has a kinetic, spatial fluctuation and potential energy term. Then, the following expression is used as an energy source term in the numerical simulation
\begin{equation}
S^0_{\eta+\xi}=\frac{1}{\Delta t}\left(\Delta E_\eta+\Delta E_\xi\right)\, .
\end{equation}
In \cite{Nahrgang:2011ll} we studied the relaxation of the sigma field in a static heat bath including this source term to account for the energy exchange. Here we present calculations for the expanding fluid where the momentum conservation must be taken into account, too. The spatial part of the source term, which is responsible for the momentum exchange, is calculated in a similar way. The momentum transfer due to the interaction and dissipation is given by
\begin{equation}
\Delta\vec{M}_\eta=\left(g\rho_s(x)+\eta(x)\partial_t\sigma(x)\right)\vec{\partial}\sigma(x)\Delta t\, ,
\end{equation}
and $\Delta\vec{M}_\xi$ is obtained from the comparision to the change in the momentum density of the field
\begin{equation}
\vec{M}_\sigma=\partial_t\sigma\vec{\partial}\sigma\, .
\end{equation}
Then, the spatial part of the source term reads
\begin{equation}
\vec{S}_{\eta+\xi}=\frac{1}{\Delta t}\left(\Delta \vec{M}_\eta+\Delta \vec{M}_\xi\right)\, .
\end{equation}

On the fluid dynamic side of the coupled evolution we apply the approximation of an ideal fluid by neglecting any corrections to the energy-momentum tensor of the quarks given by equation (\ref{hic_idealfluidtmunu}).

\begin{figure}
 \begin{center}
 \subfigure[]{\label{fig:cf2_totenergyetaTfo}\includegraphics[width=0.45\textwidth]{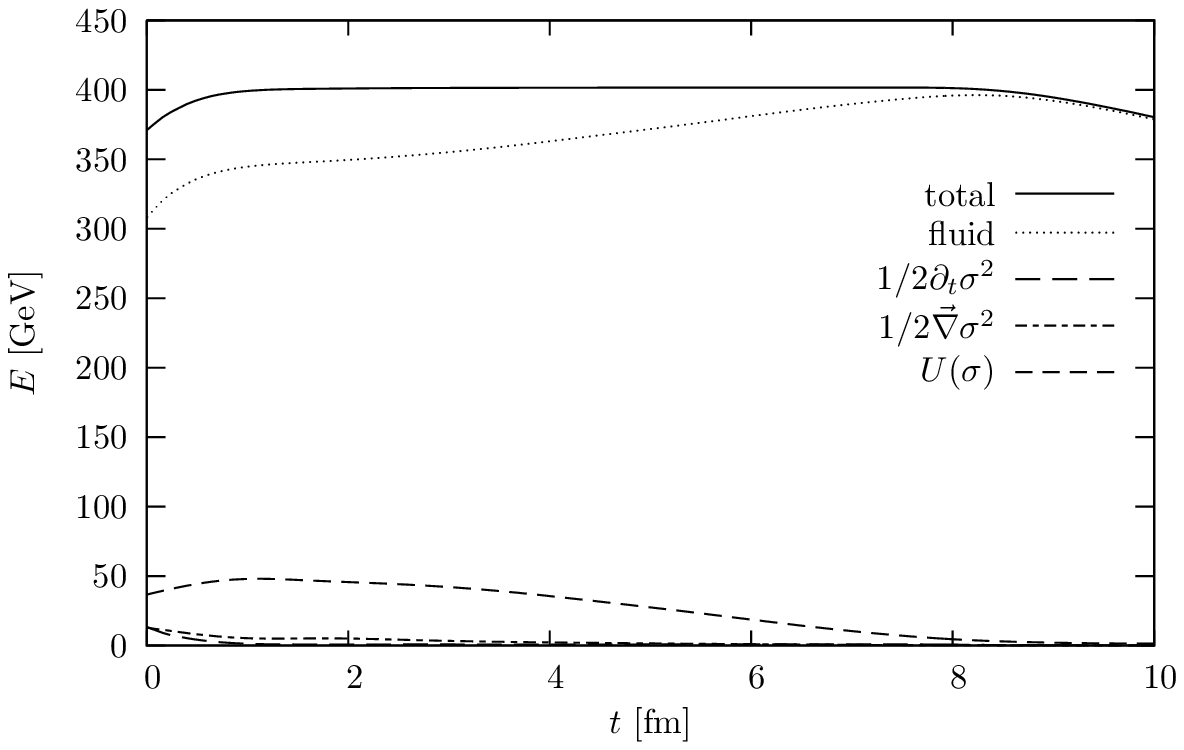}}
\subfigure[]{\label{fig:cf2_totenergyetaTcp}\includegraphics[width=0.45\textwidth]{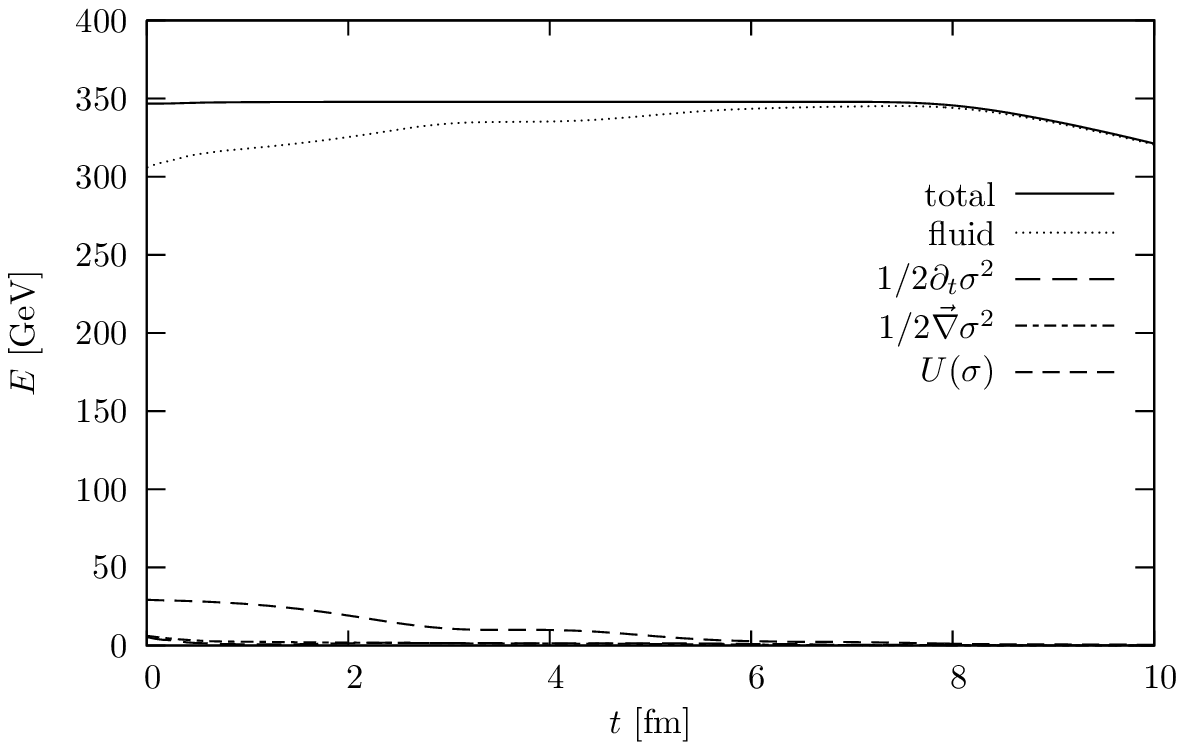}}
\end{center}
\caption{The total energy of the system for a scenario with a first order phase transition \subref{fig:cf2_totenergyetaTfo} and with a critical point \subref{fig:cf2_totenergyetaTcp}. It is the sum of the fluid energy in the laboratory frame and the field energy given in equation (\ref{eq:cf_etotsigma}).}
 \label{fig:cf2_totenergyetaT}
\end{figure}

In the beginning of the simulation the field configuration gains energy due to increasing fluctuations. In some cells this increase in field energy may exceed the energy in the fluid. In these cells the energy density is set to zero at the end of each numerical time step. In the beginning we, thus, observe a net increase of the total energy by $\lesssim10$\% in the scenario with a first order phase transition, while the energy conservation is well fulfilled during the expansion.  In a scenario with a critical point the energy is very well conserved during the entire evolution. The quark fluid reaches the edges of the grid at around $t\simeq8$ fm and disappears. In a test case with a larger grid the total energy was conserved well for a longer time. In addition, the evolution of the system in the inner region was not altered. The results are shown in figure \ref{fig:cf2_totenergyetaT}.

In addition to the energy conservation we also checked the momentum conservation. The total momentum of the system is close to zero. Due to the finite initial time derivative of the sigma field each initial field configuration has a small overall momentum, which is slightly enhanced in the beginning of the simulation. The total momentum in each of the three directions is of the order of $0.1$\% of the total momentum in positive direction.

\section{Nonequilibrium effects in chiral phase transitions}\label{sec:cf2_supercooling}
With a constant damping coefficient $\eta$ one can generally investigates dissipation and relaxation \cite{Nahrgang:2010fz}.
In this section we investigate the evolution of the entire system under the  temperature-dependent damping coefficient $\eta=\eta(T)$. The temperature dependence of $\eta=\eta(T)$ includes more aspects of the phase transition scenario than a constant damping coefficient. 

\subsection{Time evolution of the coupled system}
\begin{figure}
 \centering
 \subfigure[]{\label{fig:energydenscp}\includegraphics[width=0.45\textwidth]{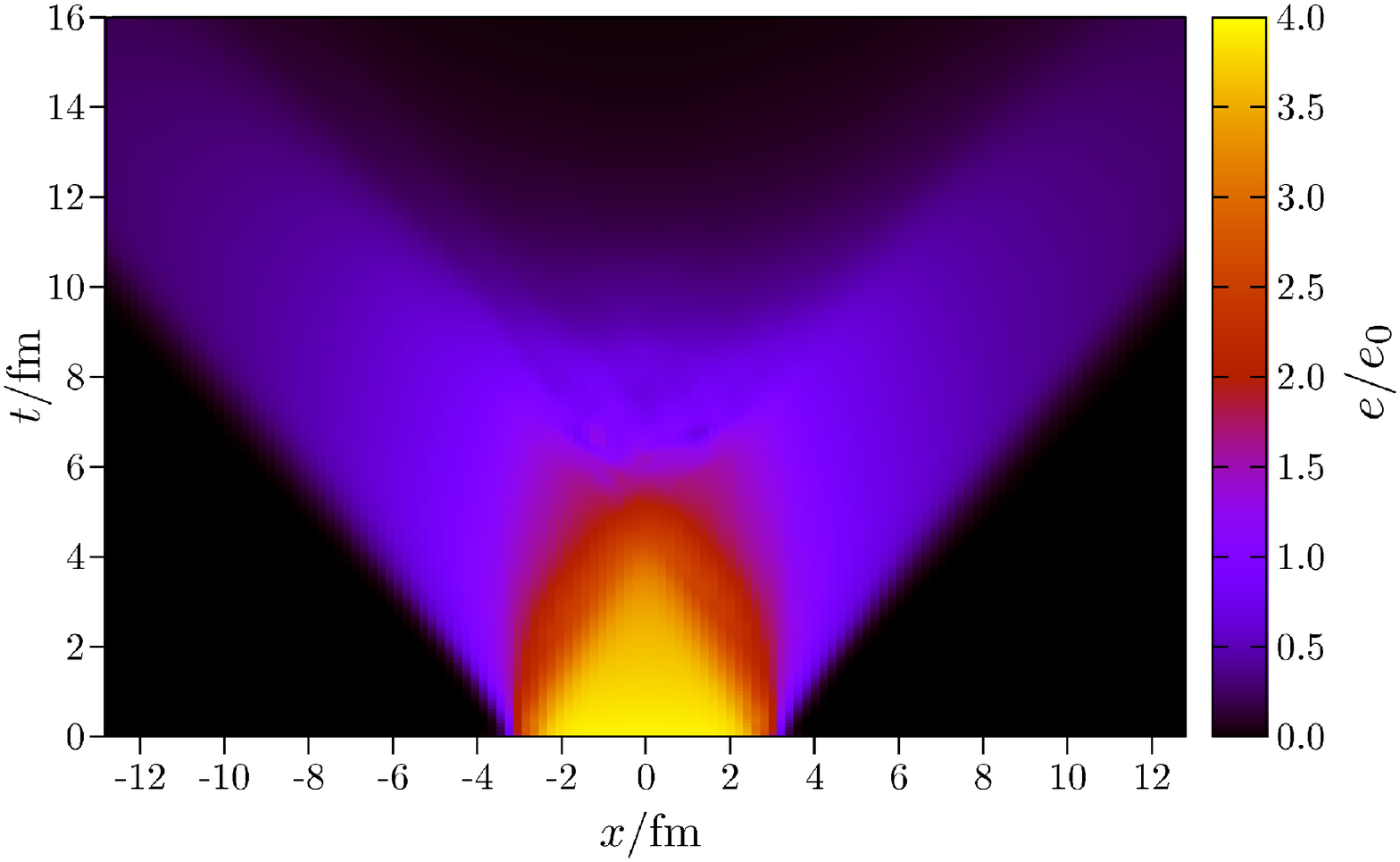}}
 \subfigure[]{\label{fig:energydensfo}\includegraphics[width=0.45\textwidth]{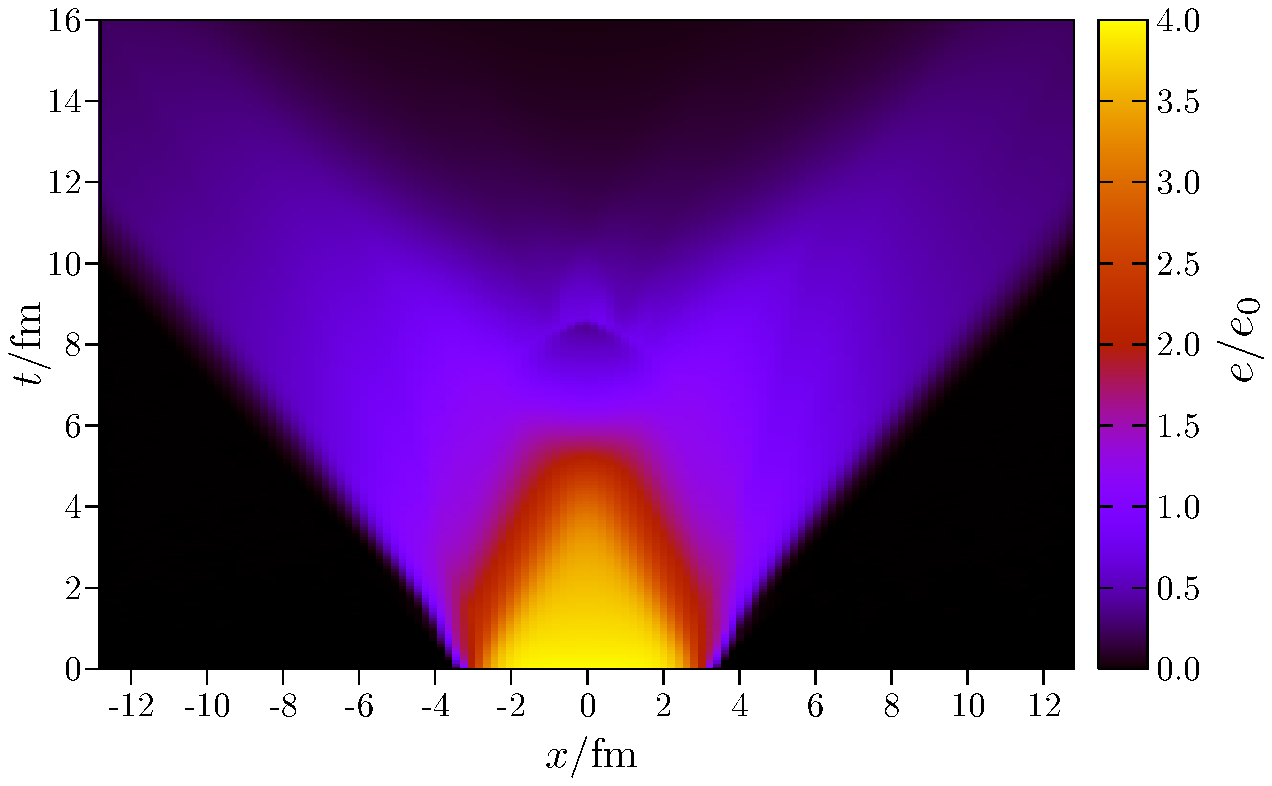}}
\caption{Time evolution of the fluid energy density along the $x$-direction in the plane $y=z=0$ for a scenario with a critical point \subref{fig:energydenscp} and with a first order phase transition in \subref{fig:energydensfo}.}
 \label{fig:energydens}
\end{figure}

\begin{figure}
 \centering
 \subfigure[]{\label{fig:sigmafieldcp}\includegraphics[width=0.45\textwidth]{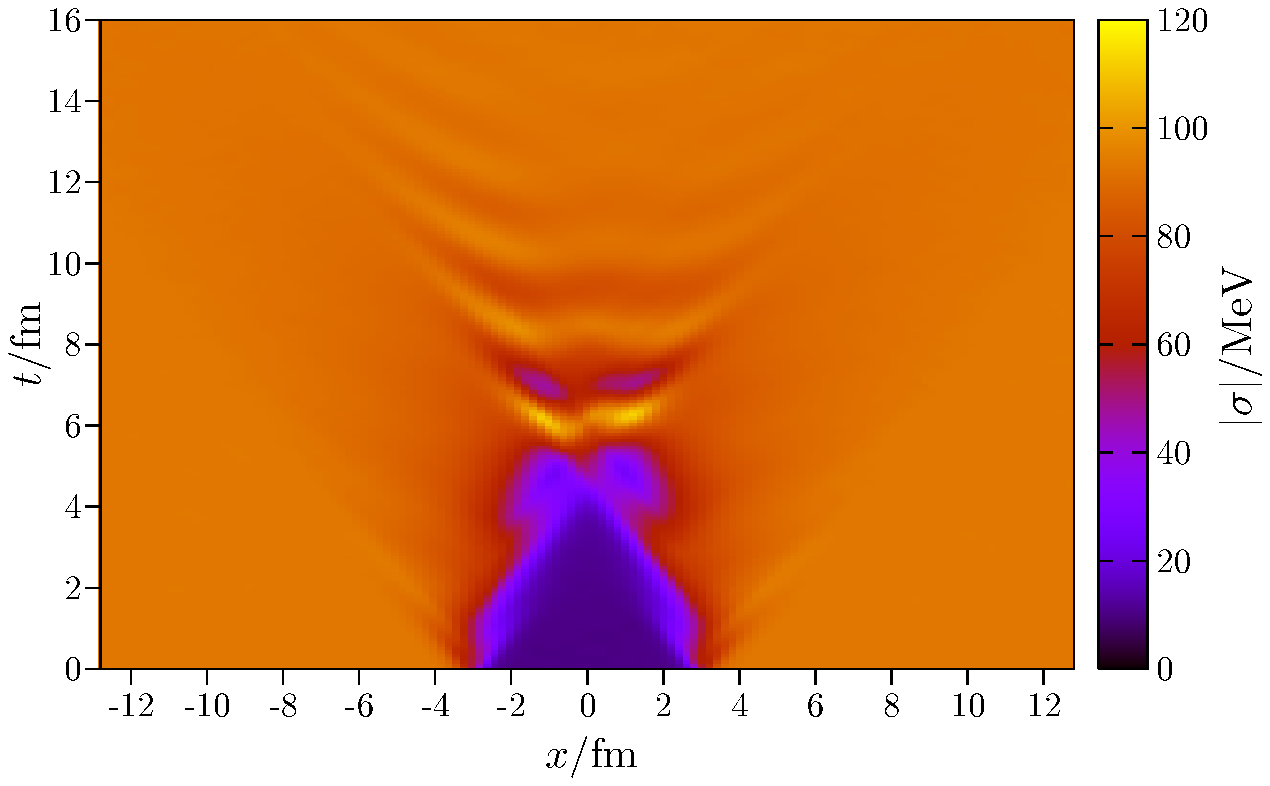}}
 \subfigure[]{\label{fig:sigmafieldfo}\includegraphics[width=0.45\textwidth]{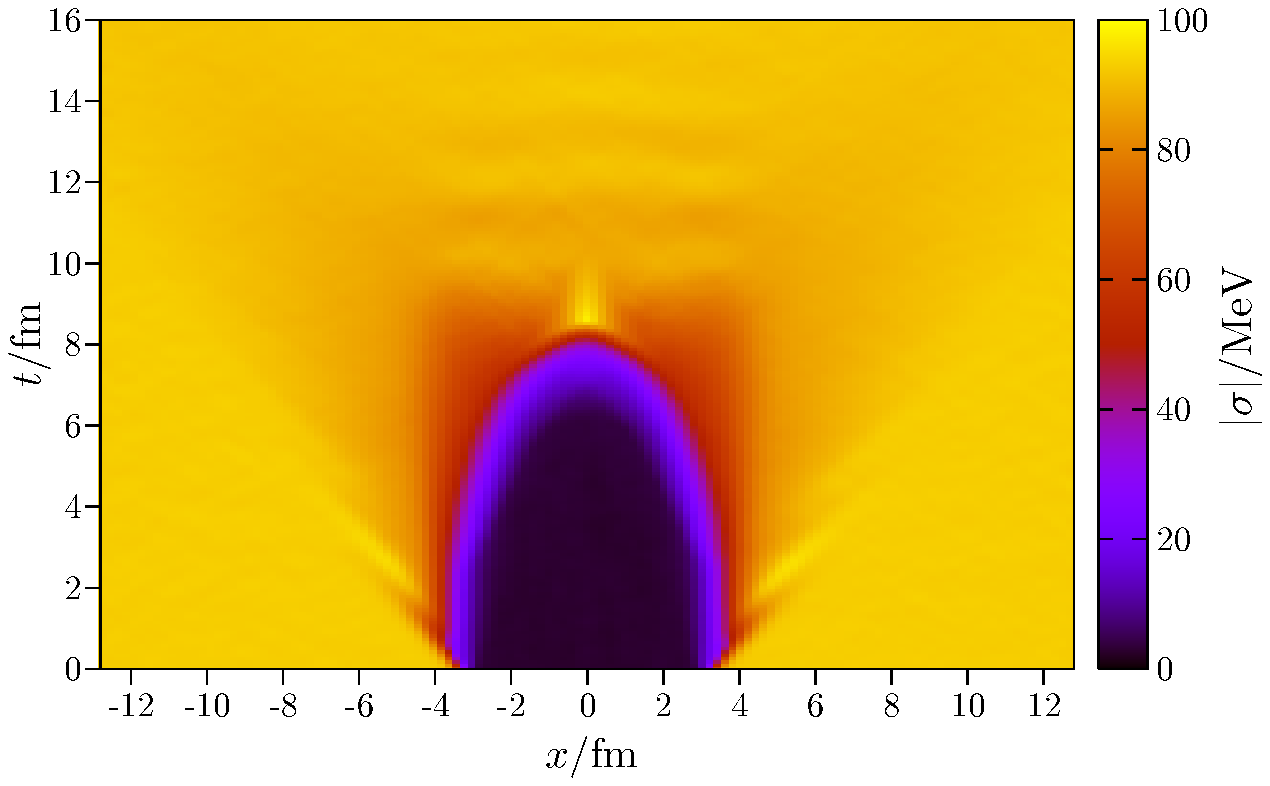}}
\caption{Time evolution of the sigma field along the $x$-direction in the plane $y=z=0$ for a scenario with a critical point \subref{fig:sigmafieldcp} and with a first order phase transition in \subref{fig:sigmafieldfo}.}
 \label{fig:sigmafield}
\end{figure}

\begin{figure}
 \centering
 \subfigure[]{\label{fig:energydenscpz}\includegraphics[width=0.45\textwidth]{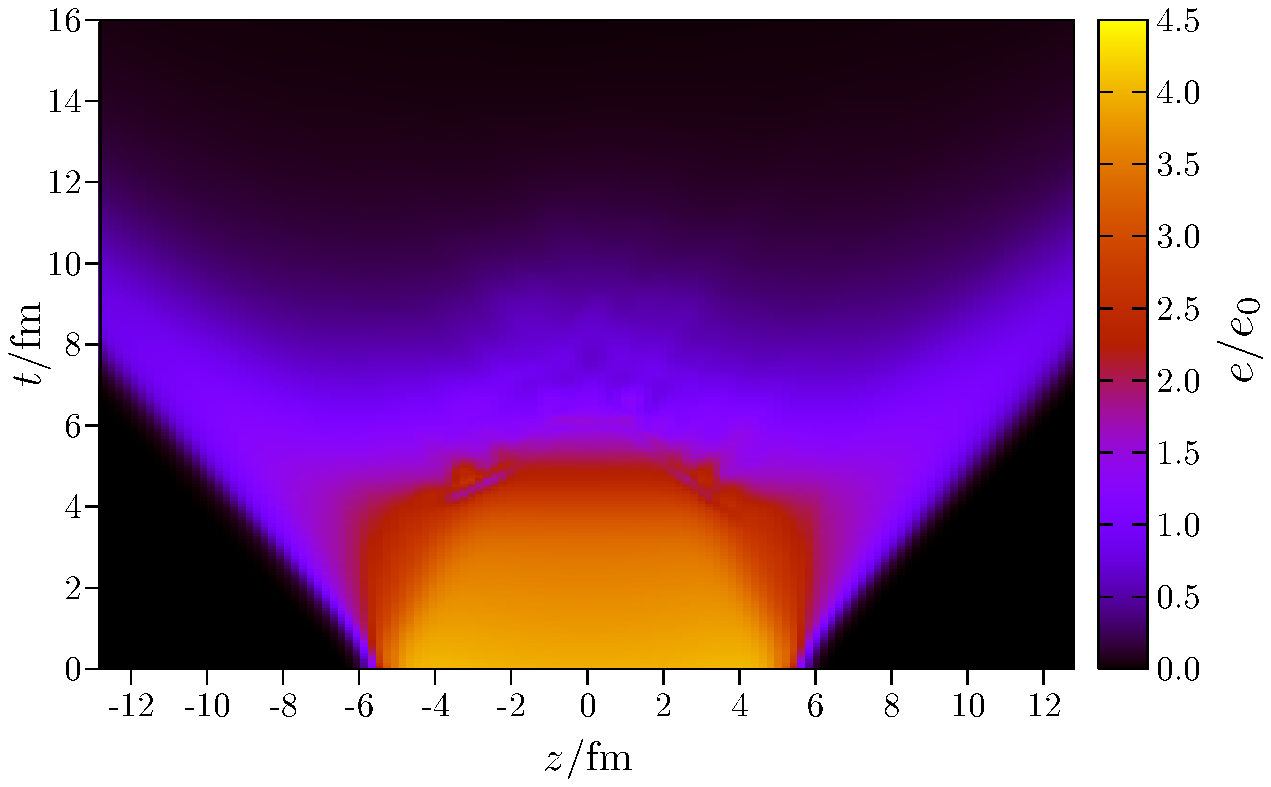}}
 \subfigure[]{\label{fig:energydensfoz}\includegraphics[width=0.45\textwidth]{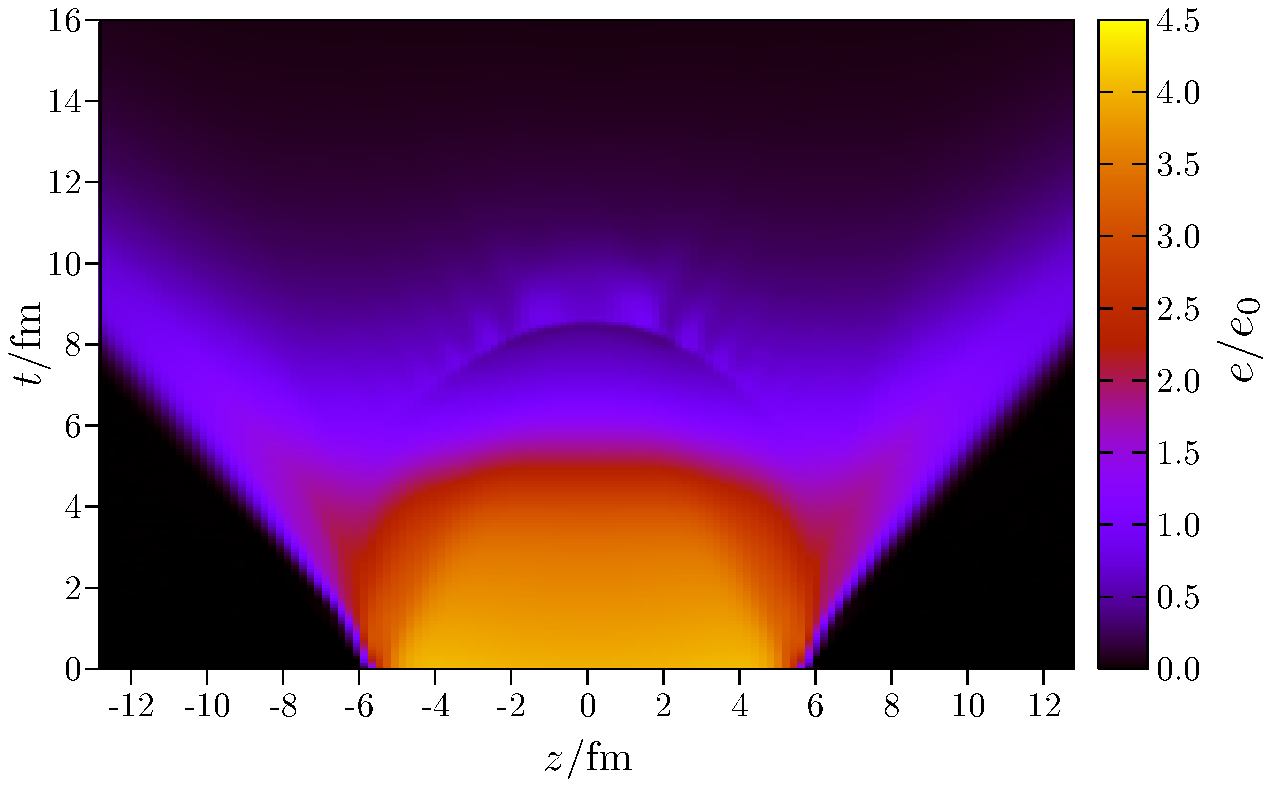}}
\caption{Time evolution of the fluid energy density along the $z$-direction in the plane $x=y=0$ for a scenario with a critical point \subref{fig:energydenscpz} and with a first order phase transition in \subref{fig:energydensfoz}.}
 \label{fig:energydensz}
\end{figure}

\begin{figure}
 \centering
 \subfigure[]{\label{fig:sigmafieldcpz}\includegraphics[width=0.45\textwidth]{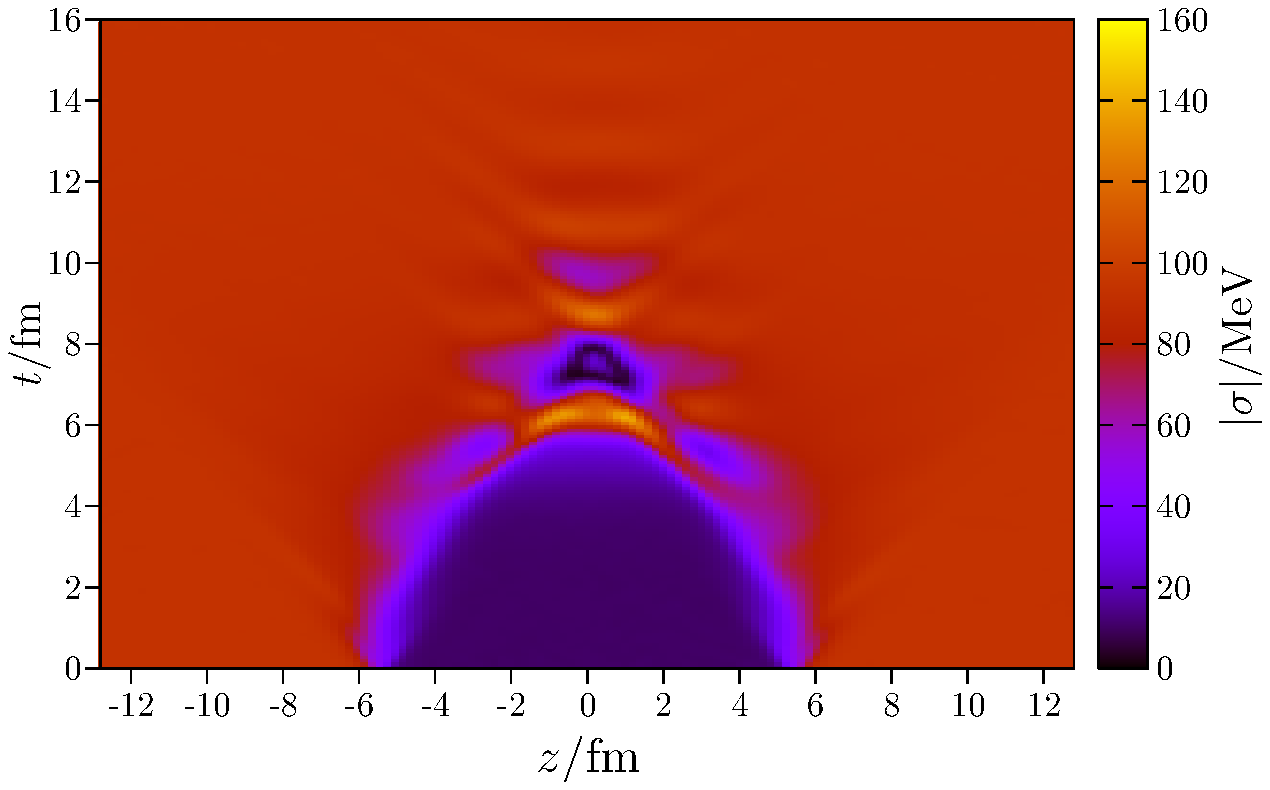}}
 \subfigure[]{\label{fig:sigmafieldfoz}\includegraphics[width=0.45\textwidth]{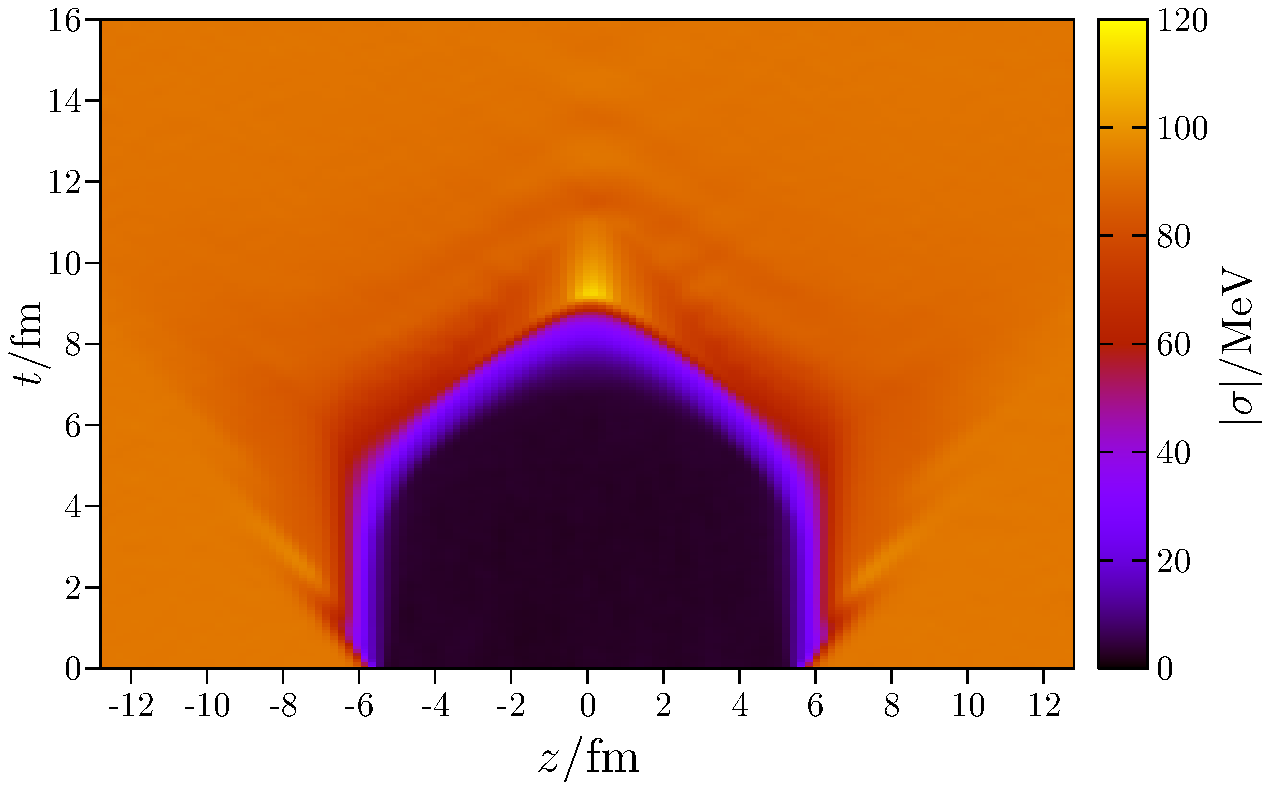}}
\caption{Time evolution of the sigma field along the $z$-direction in the plane $x=y=0$ for a scenario with a critical point \subref{fig:sigmafieldcpz} and with a first order phase transition in \subref{fig:sigmafieldfoz}.}
 \label{fig:sigmafieldz}
\end{figure}

The time evolution of the energy density and the sigma field in $x$-direction and at $y=z=0$ is shown in figures \ref{fig:energydens} and \ref{fig:sigmafield} for both phase transition scenarios. In figures \ref{fig:energydensz} and \ref{fig:sigmafieldz} we additionally show the longitudinal expansion along the $z$-direction and at $x=y=0$. We see how the quark fluid expands and the system dilutes. For a first order phase transition we can observe the energy transfer from the field to the fluid in figure \ref{fig:energydensfo}. At $x\simeq0$ and $t\simeq9$fm the sigma field finally relaxes, see figure \ref{fig:sigmafieldfo}. At the same location and place we observe an increase in the fluid energy density in figure \ref{fig:energydensfo}. This effect is less pronounced in a critical point scenario. Here, relaxation of the sigma field occurs a lot faster, see figure  \ref{fig:energydenscp}, and shows a qualitatively different behavior than for a scenario with a first order phase transition. During the relaxational process there are oscillations for the critical point scenario, because the damping coefficient vanishes for a certain region around the phase transition.

\begin{figure}
  \centering
  \includegraphics[width=0.45\textwidth]{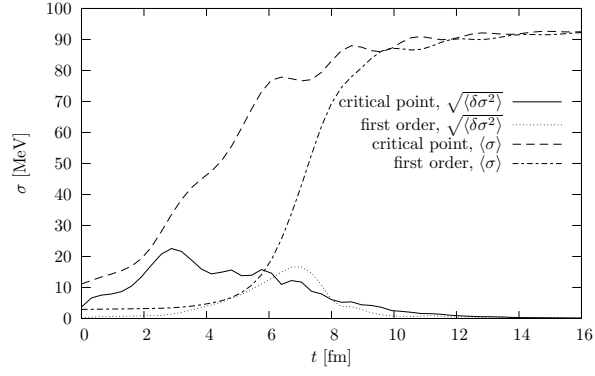}
\caption[Average and variance of the sigma field, $\eta=\eta(T)$.]{The average values and the variances of the fluctuations $\sqrt{\langle\delta\sigma^2\rangle}$ of the sigma field for $\eta=\eta(T)$ and both phase transition scenarios.}
  \label{fig:cf2_hotregionsigmaetaT}
 \end{figure}

\begin{figure}
  \centering
  \includegraphics[width=0.45\textwidth]{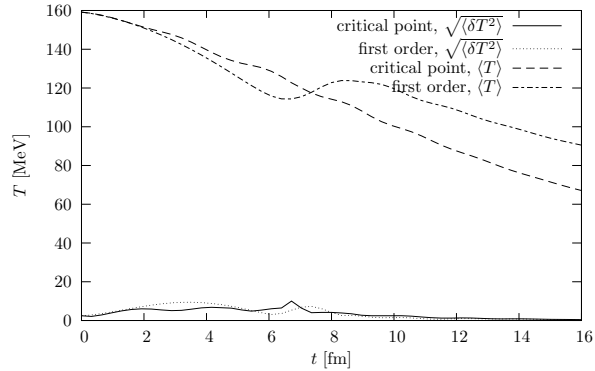}
\caption[Average and variance of the temperature, $\eta=\eta(T)$.]{The average values and the variances of the fluctuations $\sqrt{\langle \delta T^2\rangle}$ of the temperature for $\eta=\eta(T)$ and both phase transition scenarios.}
  \label{fig:cf2_hotregiontempetaT}
 \end{figure}

The time evolution of the average value of the sigma field $\langle\sigma\rangle$ and its fluctuations $\sqrt{\langle\delta\sigma^2\rangle}=\sqrt{\langle(\sigma-\langle\sigma\rangle)^2\rangle}$ are shown in figure \ref{fig:cf2_hotregionsigmaetaT}. Figure \ref{fig:cf2_hotregiontempetaT} shows the time evolution of the average temperature $\langle T\rangle$ and its fluctuations $\sqrt{\langle\delta T^2\rangle}=\sqrt{\langle(T-\langle T\rangle)^2\rangle}$. The average is taken over an initially hot and dense sphere with radius $r=3$~fm. We first discuss the critical point scenario and then the scenario  with a first order phase transition.

 The phase transition temperature in a critical point scenario $T_c=139.88$~MeV is crossed at around $t=4$~fm. Fluctuations of the sigma field are largest between  $t=2$~fm and $t=4$~fm and stay enhanced until around $t=10$~fm. The average sigma field smoothly relaxes towards its vacuum value. Strong oscillations of the average sigma field occur during the relaxational process, when large parts of the system do not experience damping. The effective potential with a critical point is very flat at the transition temperature and reheating is not observed. Instead the cooling is slightly decelerated between $t=5$~fm and $t=6$~fm as seen in figure \ref{fig:cf2_hotregiontempetaT}.

There are two reasons why the average sigma field in the scenario with a first order phase transition relaxes at later times than for a critical point scenario. First, the phase transition temperature of the first order phase transition, $T_c=123.27$~MeV, is lower than at a critical point. 
And second, the sigma field is more strongly damped at high temperatures in a scenario with a first order phase transition than in the critical point scenario. At the phase transition temperature $T_c$ the sigma mass drops below the threshold of quark-antiquark production but is larger than twice the pion mass. The damping coefficient shows a discontinuity from $\eta(T>Tc)\simeq 26.6$/fm to  $\eta(T<Tc)= 2.2$/fm. In this scenario the average of the sigma field stays constant up to almost $t\simeq 5$~fm and the relaxational process starts only when the average temperature is below the phase transition temperature as can be seen by comparing figure \ref{fig:cf2_hotregionsigmaetaT} with \ref{fig:cf2_hotregiontempetaT}. The cooling of the system is inhomogeneous. While the center is still very hot, outer parts have cooled down already. The fluctuations of temperature within the inner region with radius $r=3$~fm have a variance of almost $10$~MeV. We conclude that even in this small inner region the outer parts have already cooled below the phase transition temperatures at $t=5$~fm and that these parts relax first due to the lower damping coefficient $\eta$. When large parts of the system experience the smaller damping of $\eta=2.2$/fm the average sigma field shows oscillations during the final relaxational process also for a scenario with a first order phase transition.

\subsection{Supercooling and Reheating}
When the first order phase transition temperature is reached after $t\simeq 5$~fm large parts of the system are still in the chirally symmetric phase as the average value of the sigma field is still $\langle\sigma\rangle\lesssim10$~MeV. These large deviations of the sigma field from its equilibrium value is the nonequilibrium effect of supercooling. Due to the barrier separating the two minima of the thermodynamic potential, the sigma field is traped in the chirally symmetric state even at $T<T_c$, until the barrier disappears. Then the sigma field rolls down into a lower minimum corresponding to the broken phase.

 The released potential energy is transformed effectively into kinetic energy of field oscillations, which leads to the dissipation of energy into the fluid via $\eta(\partial_t\sigma)^2$ in the source term (\ref{eq:sc_sourceterm}). In figure \ref{fig:cf2_hotregiontempetaT} we can clearly observe the reheating effect at the first order phase transition. Between $t=7$~fm and $t=9$~fm the system in the central region is reheated from $T\simeq 114$~MeV below $T_c$ to $T\simeq 124$~MeV slightly above $T_c$, followed by a subsequent cooling. Thus, the reheating causes the system to cross the phase transition boundary two more times, once in the reverse direction from the low temperature phase to the high temperature phase around $t=8$~fm and again from above to below at around  $t=9$~fm. This explains the delayed relaxation of the average sigma field in the scenario with a first order phase transition compared to the critical point scenario. Moreover, during the relaxation and reheating process the fluctuations in the sigma field are enhanced between $t=5$~fm and $t=8$~fm.
 
\subsection{The intensity of sigma fluctuations}\label{sec:cf2_intensitysigmaflus}
In various studies on the formation of  DCC a large amplification of pion fields was observed at the phase transition \cite{Rajagopal:1993ah,Biro:1997va,Mishustin:1998yc,Xu:1999aq}. The pionic excitations were found to be triggered by violent oscillations of the sigma modes. Moreover, the zero mode of the sigma field is the order parameter of chiral symmetry breaking. The soft modes are thus expected to show large fluctuations at a critical point in thermodynamic systems. We are, therefore, especially interested in the intensity of sigma fluctuations at the phase transition.

The intensity of the sigma fluctuations is given by \cite{Abada:1996bw,AmelinoCamelia:1997in}
\begin{equation}
 \frac{{\rm d}N_\sigma}{{\rm d}^3k}=\frac{a_k^* a_k}{(2\pi)^3 2\omega_k}=\frac{1}{(2\pi)^3 2\omega_k}\bigl(\omega_k^2|\delta\sigma_k|^2+|\partial_t\sigma_k|^2\bigl)\, ,
\label{eq:cf2_numbersigma}
\end{equation}
where $a_k^*$ and $a_k$ are the Fourier coefficients in the expansion of the sigma fluctuations around the equilibrium value, $\delta\sigma(x)=\sigma(x)-\sigma_{\rm eq}(x)$, and its conjugate momentum field $\partial_t\sigma(x)$.

At later times when the nonlinearities in the equation of motion can be neglected the quantity $N_\sigma$ in (\ref{eq:cf2_numbersigma}) gives the number of sigma particles produced from the excitations of the sigma field. 
The energy $\omega_k$ of the $k$th mode of the sigma field is given by
\begin{equation}
 \omega_k=\sqrt{|k|^2+m_{\sigma}^2}\, .
\end{equation}
We define the sigma mass via the curvature of the effective potential at its equilibrium value corresponding to the average temperature $T_{\rm av}$ in a hot region of radius $r=3$~fm. It turns out that the intensity of sigma fluctuations does not depend strongly on the choice of the radius, as long as it does not extend too far into the surrounding vacuum.

At the end of each time step the Fourier transform of the sigma field is calculated by a Fast Fourier Transform algorithm. These algorithms are implemented effectively only on lattices with $N=2^n$, $n\in\mathbb{N}$, cells in each direction. This limits the realistic choice of how to define the entire volume of the system. We perform Fourier transformations on the entire grid. 

\begin{figure}
 \centering
 \subfigure[]{\label{fig:cf2_numersigmaetaT_cp}\includegraphics[width=0.45\textwidth]{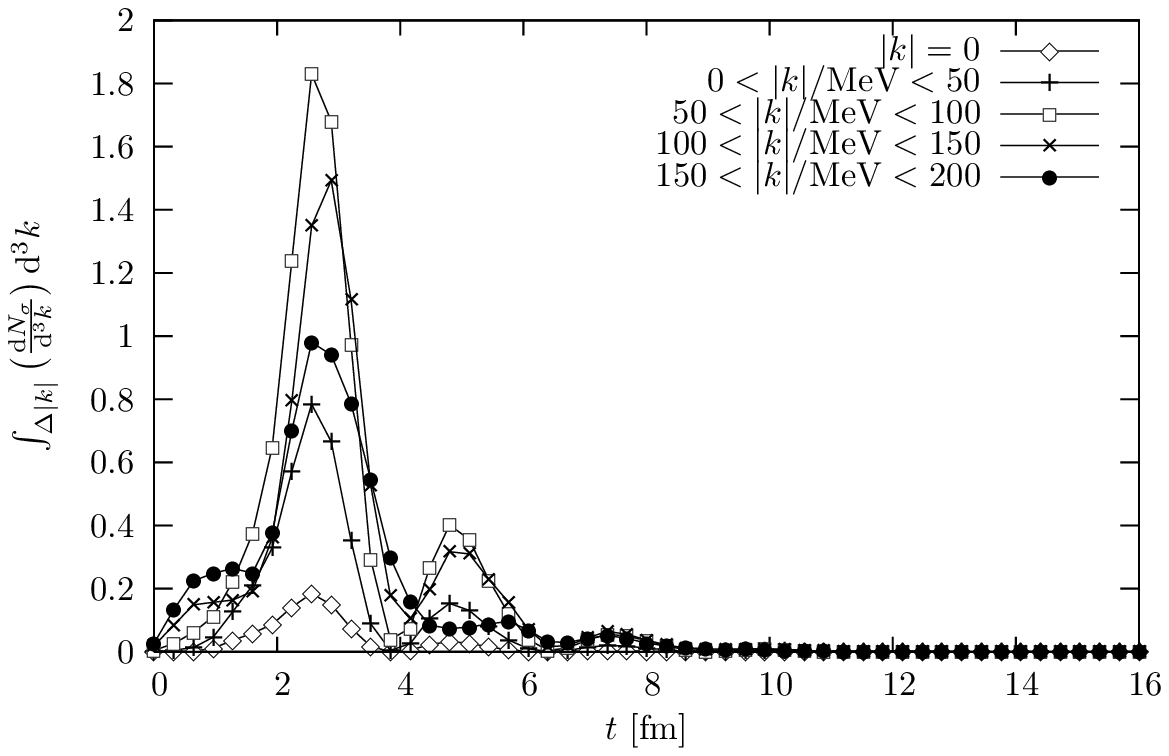}}
 \subfigure[]{\label{fig:cf2_numersigmaetaT_fo}\includegraphics[width=0.45\textwidth]{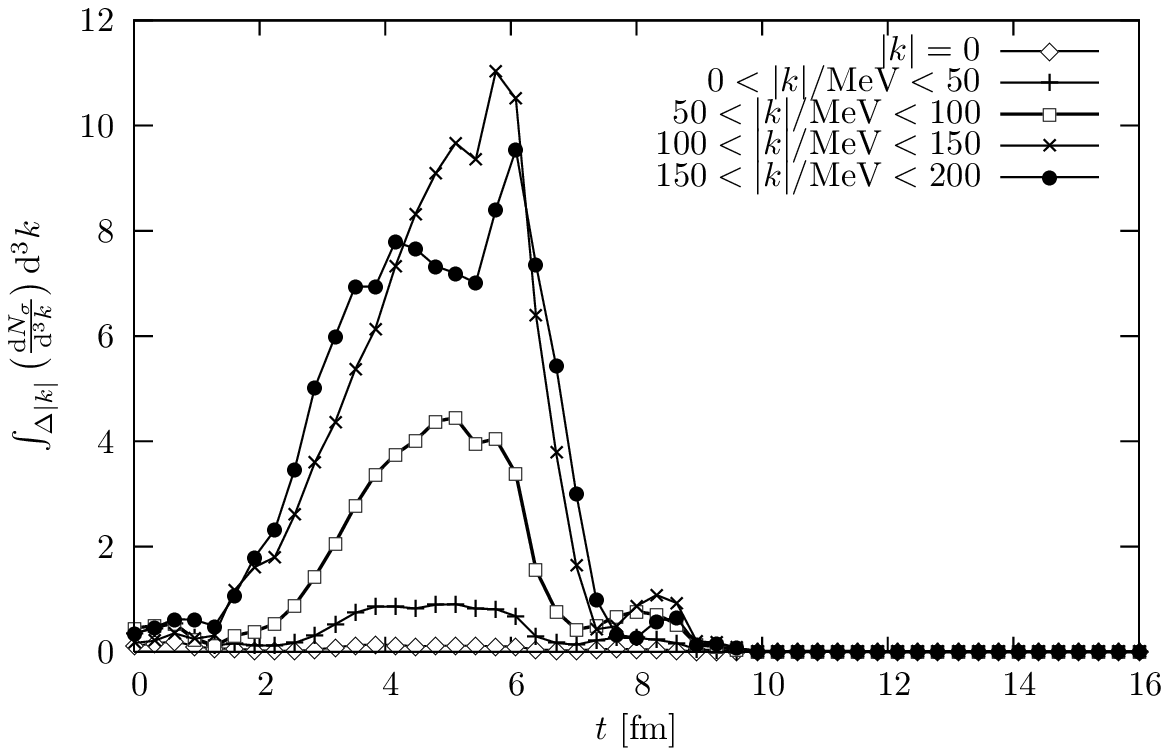}}
 \caption{The time evolution of the intensity of sigma fluctuations for a critical point scenario \subref{fig:cf2_numersigmaetaT_cp} and for a scenario with a first order phase transition \subref{fig:cf2_numersigmaetaT_fo}.}
 \label{fig:cf2_numersigmaetaT_tot}
\end{figure}

 \begin{figure}
  \centering
  \includegraphics[width=0.45\textwidth]{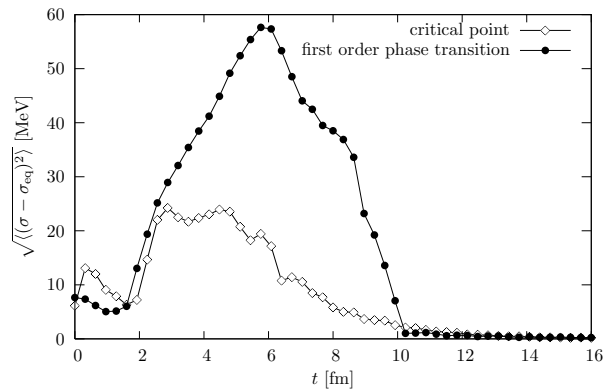}
  \caption{The time evolution of the deviation of the sigma field from its equilibrium value for a critical point scenario and a scenario with a first order phase transition.}
  \label{fig:cf2_flucgleichgewicht_beide}
 \end{figure}

For the two phase transition scenarios we show the intensity of sigma fluctuations for the zero mode and low momentum modes in momentum bins of $\Delta|k|=50$~MeV. Finally, we compare these values to the development of the deviations of the sigma field from its thermal equilibrium value averaged over the initially hot and dense region. The initial conditions, see section \ref{sec:cf1_inicond}, are chosen such that the sigma field is in equilibrium with the quark fluid at the initial temperature distributed according to  (\ref{eq:lsm_iniT}). In both phase transition scenarios the intensity of sigma fluctuations is increased during the expansion and cooling see figure \ref{fig:cf2_numersigmaetaT_tot}. For a critical point scenario the results are shown in figure \ref{fig:cf2_numersigmaetaT_cp}. The intensity shows a peak at $t=2$~fm and a second but smaller peak at $t=6$~fm. Between $t=2$~fm and $t=6$~fm the deviations of the sigma field from its equilibrium value are enhanced, see figure \ref{fig:cf2_flucgleichgewicht_beide}.

For the scenario with a first order phase transition the time evolution of the intensity of sigma fluctuations is shown in figure \ref{fig:cf2_numersigmaetaT_fo}. Figure \ref{fig:cf2_flucgleichgewicht_beide} shows the deviations of the sigma field from its equilibrium value. Since these deviations are larger than for the scenario with a critical point, the intensity of sigma fluctuations is also larger. 
Largest deviations are seen around $t=6$~fm, where large parts of the system are supercooled. Then, the system quickly relaxes to a new ground state. This process is accompanied by reheating of the heat bath. This explains a shoulder, before the system relaxes and the deviations vanish.

\section{Summary and Outlook}\label{sec:summ}
We presented a consistent dynamic model to describe nonequilibrium phase transitions in heavy-ion collisions. We included damping and noise terms originating from the interaction of the chiral fields with the quark fluid. To ensure energy and momentum conservation of the coupled system we introduced a source term into the equations of relativistic fluid dynamics and thus explicitly took the back reaction on the heat bath into account. The various aspects of energy and momentum conservation were discussed in detail. 
We investigated the expansion of the coupled system with a damping coefficient $\eta=\eta(T)$ for a scenario with a critical point and with a first order phase transition. Nonequilibrium effects like supercooling and reheating are clearly observed in case of the first order phase transition. The number of coherently produced soft sigmas is much larger in a scenario with a first order phase transition than in a scenario with a critical point. 

Several extensions of the presented framework are conceivable which would make the approach quantitatively more realistic. Most straightforward is the inclusion of soft pions as classical fields subject to stochastic dynamics and of hard pions which contribute to the heat bath. 
Possible extensions include gluonic degrees of freedom in the form of a dilaton field  \cite{Mishustin:1998yc} or the Polyakov loop \cite{Schaefer:2007pw}.
 This leads presumably to a more realistic phase diagram on the temperature-chemical potential plane. Finally a finite baryo-chemical potential can be introduced which allows to access different types of the phase transition without varying the coupling constant. The comparison of the fluctuation dynamics in the static and dynamical background is a particularly interesting topic for future studies \cite{workinprogress}.

 The authors thank Carsten Greiner for fruitful discussions and Dirk Rischke for providing the SHASTA code. M.~Nahrgang acknowledges financial support from the Stiftung Polytechnische Gesellschaft Frankfurt.
 This work was supported by the Hessian LOEWE initiative Helmholtz International Center for FAIR.


\begin{thebibliography}{50}

\bibitem{Stephanov:1998dy}
  M.~A.~Stephanov, K.~Rajagopal and E.~V.~Shuryak,
  Phys.\ Rev.\ Lett.\  {\bf 81} (1998) 4816

\bibitem{Stephanov:1999zu}
  M.~A.~Stephanov, K.~Rajagopal and E.~V.~Shuryak,
  Phys.\ Rev.\  D {\bf 60} (1999) 114028
\bibitem{Mishustin:1998eq}
  I.~N.~Mishustin,
  Phys.\ Rev.\ Lett.\  {\bf 82} (1999) 4779

\bibitem{Randrup:2010ax}
  J.~Randrup,
  Phys.\ Rev.\  {\bf C82 } (2010)  034902.
  [arXiv:1007.1448 [nucl-th]].

\bibitem{Caines:2009yu}
  H.~Caines [ STAR Collaboration ],
   [arXiv:0906.0305 [nucl-ex]].


\bibitem{Friman:2011zz}
  B.~Friman, C.~H\"ohne, J.~Knoll, S.~Leupold, J.~Randrup, R.~Rapp, P.~Senger (eds.),
  Lect.\ Notes Phys.\  {\bf 814 } (2011)  1-980. 

\bibitem{nica:whitepaper}
theor.jinr.ru/twiki-cgi/view/NICA/NICAWhitePaper

\bibitem{Berdnikov:1999ph}
  B.~Berdnikov and K.~Rajagopal,
  Phys.\ Rev.\  D {\bf 61}, 105017 (2000)

\bibitem{Csernai:1992tj}
  L.~P.~Csernai, J.~I.~Kapusta,
  Phys.\ Rev.\  {\bf D46 } (1992)  1379-1390.


\bibitem{Chomaz:2003dz}
  P.~Chomaz, M.~Colonna, J.~Randrup,
  Phys.\ Rept.\  {\bf 389 } (2004)  263-440.

\bibitem{Rajagopal:1993ah}
  K.~Rajagopal and F.~Wilczek,
  Nucl.\ Phys.\  B {\bf 404} (1993) 577
\bibitem{Biro:1997va}
  T.~S.~Biro and C.~Greiner,
  Phys.\ Rev.\ Lett.\  {\bf 79} (1997) 3138
\bibitem{Xu:1999aq}
  Z.~Xu and C.~Greiner,
  Phys.\ Rev.\  D {\bf 62} (2000) 036012

\bibitem{Morikawa:1986rp}
  M.~Morikawa,
  Phys.\ Rev.\  {\bf D33 } (1986)  3607.

\bibitem{Gleiser:1993ea}
  M.~Gleiser and R.~O.~Ramos,
  Phys.\ Rev.\  D {\bf 50} (1994) 2441

\bibitem{Boyanovsky:1996xx}
  D.~Boyanovsky, I.~D.~Lawrie, D.~S.~Lee,
  Phys.\ Rev.\  {\bf D54 } (1996)  4013-4028.

\bibitem{Greiner:1996dx}
  C.~Greiner and B.~Muller,
  Phys.\ Rev.\  D {\bf 55} (1997) 1026

\bibitem{Bodeker:1995pp}
  D.~Bodeker, L.~D.~McLerran, A.~V.~Smilga,
  Phys.\ Rev.\  {\bf D52 } (1995)  4675-4690.

\bibitem{Son:1997qj}
  D.~T.~Son,
  [hep-ph/9707351].

\bibitem{Rischke:1998qy}
  D.~H.~Rischke,
  Phys.\ Rev.\  C {\bf 58} (1998) 2331

\bibitem{Fraga:2004hp}
  E.~S.~Fraga and G.~Krein,
  Phys.\ Lett.\  B {\bf 614} (2005) 181


\bibitem{Mishustin:1998yc}
  I.~N.~Mishustin and O.~Scavenius,
  Phys.\ Rev.\ Lett.\  {\bf 83} (1999) 3134

 \bibitem{Csernai:1995zn}
   L.~P.~Csernai, I.~N.~Mishustin,
   Phys.\ Rev.\ Lett.\  {\bf 74 } (1995)  5005-5008.

\bibitem{Mishustin:1997ff}
  I.~N.~Mishustin, O.~Scavenius,
  Phys.\ Lett.\  {\bf B396 } (1997)  33-38.

\bibitem{Abada:1996bw}
  A.~Abada, M.~C.~Birse,
  Phys.\ Rev.\  {\bf D55 } (1997)  6887-6899.

  
\bibitem{Paech:2003fe}
  K.~Paech, H.~Stoecker and A.~Dumitru,
  Phys.\ Rev.\  C {\bf 68} (2003) 044907

\bibitem{workinprogress}
T.~Koide, G.~Denicol, G.~Torrieri and I.~N.~Mishustin, work in progress.

\bibitem{landaulifschitz5}
  L.~Landau, E.~Lifshitz and L~Pitaevskii, Statistical Physics, vol. I (Pergamon Press, London, 1980).

\bibitem{Nahrgang:2011mg}
  M.~Nahrgang, S.~Leupold, C.~Herold, M.~Bleicher,
  Phys.\ Rev.\  {\bf C84 } (2011)  024912.


\bibitem{Luttinger:1960ua}
  J.~M.~Luttinger, J.~C.~Ward,
  Phys.\ Rev.\  {\bf 118 } (1960)  1417-1427.

\bibitem{Lee:1960zza}
  T.~D.~Lee, C.~N.~Yang,
  Phys.\ Rev.\  {\bf 117 } (1960)  22-36.

\bibitem{Baym:1961zz}
  G.~Baym, L.~P.~Kadanoff,
  Phys.\ Rev.\  {\bf 124 } (1961)  287-299.

\bibitem{Baym:1962sx}
  G.~Baym,
  Phys.\ Rev.\  {\bf 127 } (1962)  1391-1401.


\bibitem{Cornwall:1974vz}
  J.~M.~Cornwall, R.~Jackiw and E.~Tomboulis,
  Phys.\ Rev.\  D {\bf 10} (1974) 2428.

\bibitem{Ivanov:1998nv}
  Yu.~B.~Ivanov, J.~Knoll and D.~N.~Voskresensky,
  Nucl.\ Phys.\  A {\bf 657} (1999) 413

\bibitem{vanHees:2001ik}
  H.~van Hees, J.~Knoll,
  Phys.\ Rev.\  {\bf D65 } (2002)  025010.

\bibitem{vanHees:2001pf}
  H.~van Hees, J.~Knoll,
  Phys.\ Rev.\  {\bf D65 } (2002)  105005.

\bibitem{vanHees:2002bv}
  H.~van Hees, J.~Knoll,
  Phys.\ Rev.\  {\bf D66 } (2002)  025028.



\bibitem{Scavenius:2000qd}
  O.~Scavenius, A.~Mocsy, I.~N.~Mishustin and D.~H.~Rischke,
  Phys.\ Rev.\  C {\bf 64} (2001) 045202

\bibitem{Scavenius:2000bb}
  O.~Scavenius, A.~Dumitru, E.~S.~Fraga, J.~T.~Lenaghan and A.~D.~Jackson,
  Phys.\ Rev.\  D {\bf 63} (2001) 116003

\bibitem{Aguiar:2003pp}
  C.~E.~Aguiar, E.~S.~Fraga and T.~Kodama,
  J.\ Phys.\ G {\bf 32} (2006) 179

\bibitem{Nahrgang:2011ll}
  M.~Nahrgang, S.~Leupold, M.~Bleicher,
   [arXiv:1105.1396 [nucl-th]].

\bibitem{Rischke:1995ir}
  D.~H.~Rischke, S.~Bernard, J.~A.~Maruhn,
  Nucl.\ Phys.\  {\bf A595 } (1995)  346-382.

\bibitem{Rischke:1995mt}
  D.~H.~Rischke, Y.~Pursun, J.~A.~Maruhn,
  Nucl.\ Phys.\  {\bf A595 } (1995)  383-408.

\bibitem{Brachmann:1997bq}
  J.~Brachmann, A.~Dumitru, J.~A.~Maruhn, H.~Stoecker, W.~Greiner and D.~H.~Rischke,
  Nucl.\ Phys.\  A {\bf 619} (1997) 391

\bibitem{Nahrgang:2010fz}
  M.~Nahrgang, M.~Bleicher,
    [arXiv:1011.5379 [nucl-th]].


\bibitem{AmelinoCamelia:1997in}
  G.~Amelino-Camelia, J.~D.~Bjorken, S.~E.~Larsson,
  Phys.\ Rev.\  {\bf D56 } (1997)  6942-6956.


\bibitem{Schaefer:2007pw}
  B.~-J.~Schaefer, J.~M.~Pawlowski, J.~Wambach,
  Phys.\ Rev.\  {\bf D76 } (2007)  074023.

\end{thebibliography}
\end{document}